\documentclass{optica-article}

\articletype{Research Article}

\usepackage{graphicx}
\usepackage{subcaption}
\usepackage{lineno}
\usepackage{gensymb}
\usepackage{isotope}
\usepackage{xcolor}

\usepackage{svg}

\begin{document}

\title{Two-photon rubidium clock detecting 776~nm fluorescence}

\pagestyle{fancy}
\fancyhead{}


\author{River Beard,\authormark{1} Kyle W. Martin,\authormark{1} John D. Elgin,\authormark{2} Brian L. Kasch,\authormark{2} and Sean P. Krzyzewski\authormark{2,*}}

\address{\authormark{1}BlueHalo,
 1300 Britt Street SE Albuquerque, NM 87123, USA\\
\authormark{2}Air Force Research Laboratory, 
 Space Vehicles Directorate 
 Kirtland Air Force Base 87117\\}

\email{\authormark{*}qst@afrl.af.mil}

\begin{abstract*} 
The optical atomic clock based on the $5S_{1/2} \rightarrow 5D_{5/2}$ two-photon transition in rubidium is a candidate for a next generation, manufacturable, portable clock that fits in a small size, weight, and power (SWaP) envelope. Here, we report the first two-photon rubidium clock stabilized by detecting 776~nm fluorescence. We also demonstrate the use of a multi-pixel photon counter as a low voltage substitute to a photomultiplier tube in the feedback loop to the clock laser.

\end{abstract*}

\section{Introduction}
Robust optical clocks are critical for realizing improvements in outside-of-the-laboratory applications such as tests of general relativity\cite{Herrmann_2018}, navigation, and communication \cite{Maleki_2005}. The Optical-Rubidium Atomic Frequency Standard (O-RAFS) is candidate for fulfilling this role, with short-term fractional instabilities on the order of $10^{-13} / \sqrt{\tau}$ \cite{Martin2018,Martin2019,Newman2019,Newman2021,Lemke2022} and long-term instabilities on the order of $10^{-15}$ \cite{Lemke2022}.

O-RAFS offers several advantages over other optical clocks for realizing a portable device. Maintenance of rubidium vapor at 100\degree{C} has already been demonstrated for a mm-scale cell with 44~mW of electrical heater power \cite{Maurice2020}. O-RAFS continuously probes a $\tau \approx$~230~ns transition \cite{Sheng_2008,Safronova_2004,Safronova_2011}, so rapid feedback can be provided to the laser frequency without the need for pre-stabilization. The clock transition is at 778~nm, which is both easily integrated with commercial fiber-optic components and is significantly more accessible to modern laser technology than the ultra-violet transitions relied upon in some other portable atomic clocks \cite{Park_2020,Xin_2022}. Compared to other clocks with complicated optical schemes, O-RAFS can be designed to be mechanically robust since it requires only a few optical components. Other optical clock architectures based on vapor cells have demonstrated similar, promising results. For example, a molecular iodine clock referencing a rovibronic transition at 532~nm \cite{Schuldt_2021} was demonstrated on-board a sounding rocket \cite{Doeringshoff_2019}, and multiple iodine vapor cell clocks recently demonstrated fractional frequency stability on the order of $10^{-15}$ on board a navy vessel \cite{Roslund2023}.

We previously demonstrated the use of a multi-pixel photon counter (MPPC) as a photonically integrable alternative to the traditionally used photomultiplier tubes (PMT) \cite{IFCSPoster}. The MPPC's high detection efficiency in the near-infrared allowed for stabilization of the clock using the dominant fluorescent signal at 776~nm for the first time, reducing the clock's short-term instability associated with photon shot-noise in agreement with predictions from \cite{Hassasin2023}. In this paper, we present week-long data runs of the two-photon rubidium clock comparing performance with detection wavelengths at 420~nm and at 776~nm.

\section{Fluorescence at 776~\lowercase{nm}}
\begin{figure}[htbp]
\centering
\includegraphics[width=8.3cm]{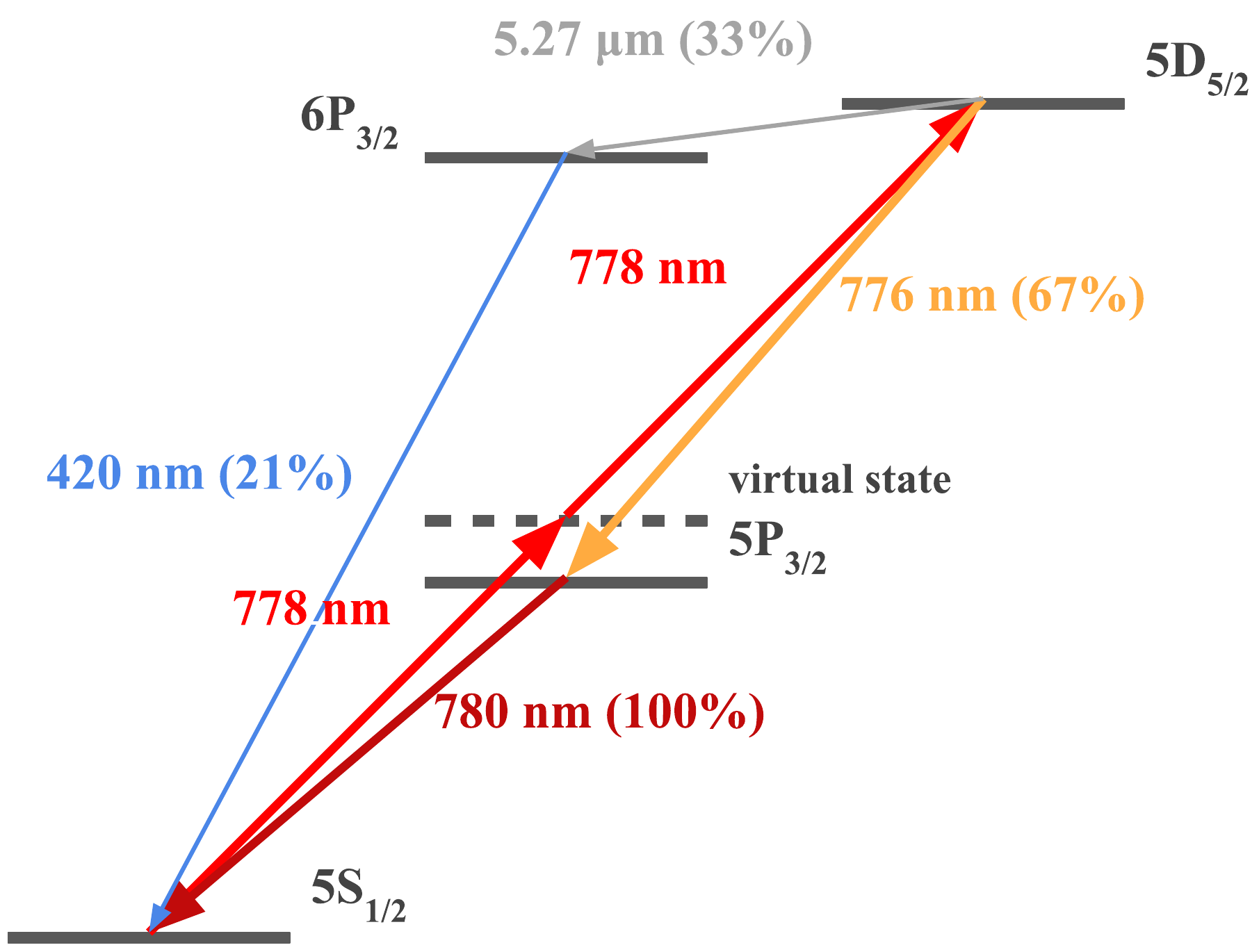}
\caption{\label{ELevels} Relevant energy levels of rubidium and branching ratios cascaded from the 5D$_{5/2}$ state using individual decay branching ratios from \cite{UDportal}. 
}
\end{figure}

The optical rubidium atomic frequency standard traditionally collects 420~nm fluorescence because it is easily filtered from the 778~nm interrogating laser. Hot-mirrors with wide collection angles and colored glass filters allow a majority of the 420~nm fluorescence to reach a low-light level detector while blocking the bright, interrogating beam. The 420~nm fluorescence is generated in a cascade decay where 33\% of excited atoms decay to the $6P_{3/2}$ state, and 21\% of atoms in the $6P_{3/2}$ state decay directly to the ground state \cite{UDportal}. Unfortunately, only 7\% of atoms excited by the interrogating laser emit at 420~nm, leading a low signal-to-noise ratio along this detection path and high short-term instability of the clock. Furthermore, some of the 420~nm fluorescence is reabsorbed by nearby ground-state atoms with the same 21\% chance of emitting 420~nm fluorescence, and is otherwise lost as infrared radiation from decays through the $6S$ or $4D$ states as described in \cite{Hassasin2023}. Detection of a stronger emission line should reduce the short-term instability of the clock. Branching ratios from \cite{UDportal} are used to calculate the percentage of atoms that contribute to each emission line. We choose to filter for the 776~nm line, which is associated with the $5D_{5/2} \rightarrow 5P_{3/2}$ decay. While slightly fewer 776~nm photons (67\%) are emitted than those at 780~nm (86\%), we choose the former because it is not connected to the ground state, and is therefore free of radiation trapping, an effect whereby the ground-state atoms absorb and isotropically re-emit the fluorescent photons, blurring the phase of the time-dependent fluorescent signal and limiting the bandwidth of the clock laser's phase-lock-loop. Relevant energy levels are shown in Fig. \ref{ELevels}.

\section{Experiment}
\subsection{Laser System}
The interrogating laser system is seeded with a fiber-coupled telecommunications band (telecom) external cavity diode laser at 1556.2~nm. The telecom laser is amplified and frequency doubled through a nonlinear crystal in a freespace cavity. The 778~nm beam passes through a freespace acousto-optic modulator (AOM) to control the optical power. The zeroeth order beam is then fiber-coupled into a polarization-maintaining optical fiber and directed toward the rubidium atomic reference, where it passes through an absorptive, linear polarizer and a non-polarizing beamsplitter before interrogating the rubidium vapor with approximately 6~mW of one-way optical power. 

Modulation of the laser current and demodulation of the MPPC's transimpedance amplifier voltage are performed digitally at 47.577~kHz, with a proportional-integral-derivative (PID) servo feedback bandwidth of 13~kHz to the direct current of the laser to stabilize the laser frequency to the atomic resonance. Coarse tuning of the laser frequency is performed by feeding back to the temperature of the telecom laser.

A self-referenced optical-fiber frequency comb is stabilized to the telecom laser, dividing the output frequency of the atomic clock down to 160~MHz.

\subsection{Rubidium Atomic Reference}

\begin{figure}[ht]
	\centering
	\caption{}\label{bothRARs}
	\begin{subfigure}{\linewidth}
    \centering
		\includegraphics[width=8.3cm]{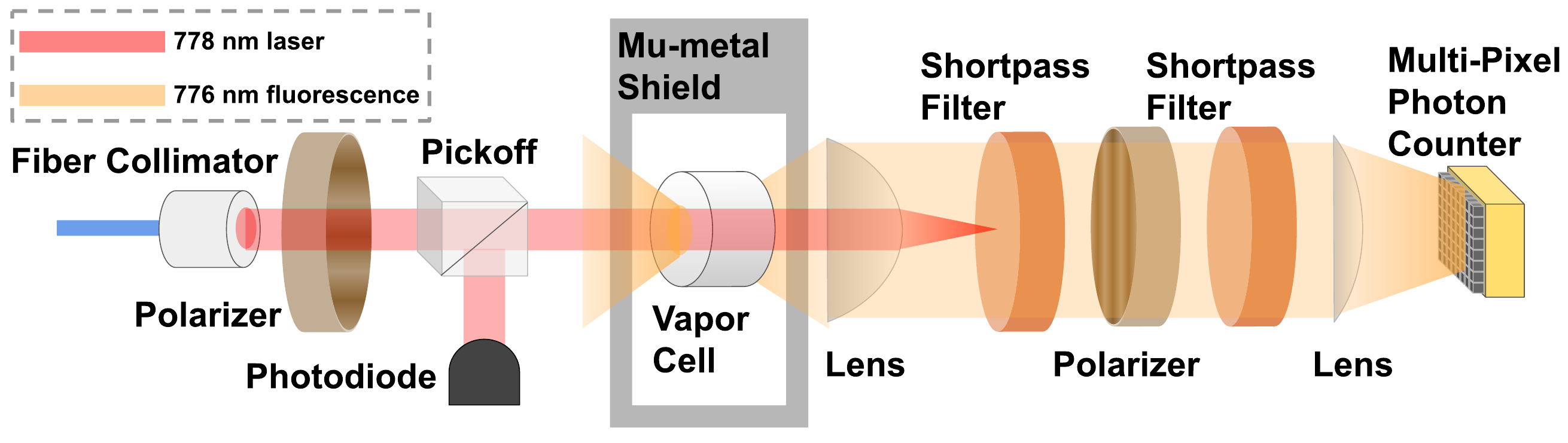}
		\caption{\label{RAR} The 778~nm clock laser is coupled out of a polarization-maintaining fiber and directed toward a mm-scale vapor cell of enriched \isotope[87]{Rb}. A high-NA lens behind the vapor cell focuses the laser onto a geometrically flat, spectrally steep short-pass filter and roughly collimates the fluorescence in the direction of the detector. The dielectric filter retroreflects the 778~nm laser light and transmits 776~nm fluorescence within 5\degree{} angle of incidence. A final lens focuses the fluorescence onto a MPPC for detection. A magnetic shield diverts external magnetic fields away from the vapor cell. High extinction-ratio cross-polarizers assist in keeping laser light off of the detector, though rejecting half of the 776~nm signal in the process. Additional steep, short-pass filters are inserted to reduce laser light on the detector. 
}
	\end{subfigure}
	\centering
	\begin{subfigure}{\linewidth}
	\centering
        \includegraphics[width=8.3cm]{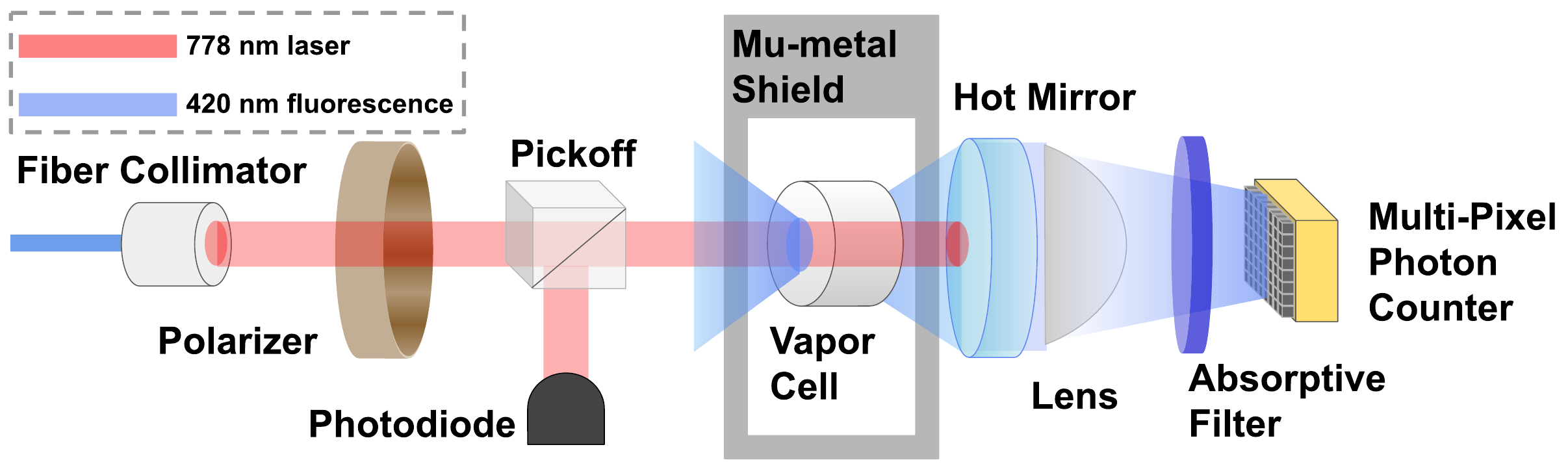}
		\caption{\label{RAR420} The same fiber collimator, initial polarizer, magnetic shield, pickoff, photodiode, and vapor cell from the 776~nm-detecting RAR are used. A hot mirror retroreflects the interrogating beam. No focusing/collimating lens precedes the retroreflector, as the hot-mirror has a large acceptance angle at 420~nm. A lens refocuses light toward the same MPPC as before. A colored-glass absorptive filter replaces the absorptive linear polarizer, and no additional short-pass filters are necessary.
}
	\end{subfigure}

\end{figure}

The Rubidium Atomic Reference (RAR) for detection at 776~nm is shown in Fig. \ref{RAR}. A RAR that detects 420~nm fluorescence was also constructed to compare the performance of the two detection schemes (Fig. \ref{RAR420}). While a millimeter-scale, blown glass cell is used in this particular experiment, the in-line optical layout presented here is compatible with microfabricated vapor cell technology. A linearly polarized beam of $1/e^2$ intensity radius $w_0 = 0.49$~mm interrogates a 5~mm-long cell of enriched \isotope[87]{Rb} vapor, whose windows are anti-reflective coated for 780~nm.

An approximately 1~mL copper oven surrounds the cylindrical vapor cell and is contacted to the cell windows with a flexible silicone elastomer to prevent rubidium from condensing in the optical path while inflicting minimal stress on the glass. The copper is heated from the outside with a low-field resistive element and rests on a ceramic pedestal within a mu-metal shield. A pair of platinum resistive temperature devices (RTDs) are epoxied directly to opposite, outer sides of the cylindrical cell wall to operate as four-point-probes, one for closed-loop temperature control to 106\degree{C} and the other as a witness. Neither RTD is in direct thermal contact with the copper block. Rubidium is observed to condense only where one or more of the RTDs are contacted to the glass, indicating that the temperature sensors determine the cold spots in this configuration.

For the 776~nm-detecting RAR, a lens and a flat, dielectric filter retroreflect the laser beam to allow for Doppler-free two-photon absorption at resonance. The dielectric filter transmits 776~nm light within $3.5\pm 1.5 \degree{}$ of normal incidence. The lens is necessary for collimating the signal before it reaches the filter because only a small portion of the isotropically emitted fluorescence begins within the filter's acceptance angle. Each dielectric filter only has about 3.5 OD of blocking at 778~nm, so additional short-pass filters and an absorptive nanoparticle linear polarizer are added behind the retroreflecting filter, bringing the residual laser light at the detector to below one-tenth of the peak fluorescence. It is conceivable that a higher quality dielectric filter could alleviate the need for the polarizer in the future, which would improve signal-to-noise ratio by 1.5~dB, since half of the fluorescence has the same polarization as the laser beam and is therefore lost in our current filtering scheme. 

For the 420~nm detecting RAR, a hot mirror retroreflects the 778~nm beam while transmitting the fluorescence at 420~nm with a large acceptance angle. An absorptive, colored glass filter suppresses residual laser light below the noise floor of the detector.

With both detection schemes, a lens focuses the fluorescence onto the detector's 7~mm$^2$ active area, though an MPPC of custom geometry would eliminate the need for this final collecting optic. Finally, a non-polarizing cube beamsplitter directs a portion on the retroreflected beam to a photodiode, which serves as the input for our 20kHz-bandwidth optical power servo. In the authors' experience, reference voltages derived from line voltages are subject to long term instabilities that couple to set-point and diode bias voltages. Therefore we operate the photodiode in photovoltaic mode to avoid the need for a stable bias voltage, and is transimpedance amplified with a resistor, though the setpoint voltage is still subject to drift.



\subsection{Multi-Pixel Photon Counter}
The two-photon clocks in \cite{Martin2018,Martin2019,Newman2019,Newman2021,Lemke2022} use a PMT to detect the fluorescence. PMTs offer high gains at low light levels, but typically require bias voltages of several hundred volts to operate. The high voltage adds an additional layer of difficulty in realizing a portable system. Multi-pixel Photon Counters (MPPCs) can be used as a low voltage, high detection-efficiency alternative to the PMT. The MPPC is in essence an array of avalanche photodiodes (pixels) operating in Geiger mode, where they are reversed biased around 5 volts beyond breakdown voltage. Each pixel serves as its own single-photon detector with a binary current output. The currents from each pixel are summed to create an output whose range of possible values is equal to the number of pixels in the array. Finally, the total photocurrent is read out as a single voltage with a transimpedance amplifier. The photonic integration of individually addressed avalanche photodiodes has already been successfully demonstrated in \cite{Setzer2021}.

For this experiment, we select an MPPC whose detection efficiency is similar at 420~nm and 776~nm (Hamamatsu C14456-3050GA) at about 16\% and 20\%, respectively, though MPPCs with higher detection efficiencies at either one of these wavelengths are commercially available. The MPPC is in a vacuum tight module and is cooled to -20\degree{C} with a thermoelectric cooler (TEC) to surpress dark counts. MPPCs can still function at room temperature, but with some additional noise. This MPPC operates at a reverse bias of 47~V and has a breakdown voltage of 42~V, much lower than the hundreds of volts typical for PMTs. The driver for the MPPC and TEC ran on a $-5$~V, ground, $+5$~V power supply and drew 0.855(5)~W of electrical power during operation, though we have not investigated designs for a new, lower SWaP driver.

\section{Results}
\begin{figure}[ht]
\centering
\includegraphics[width=10.3cm]{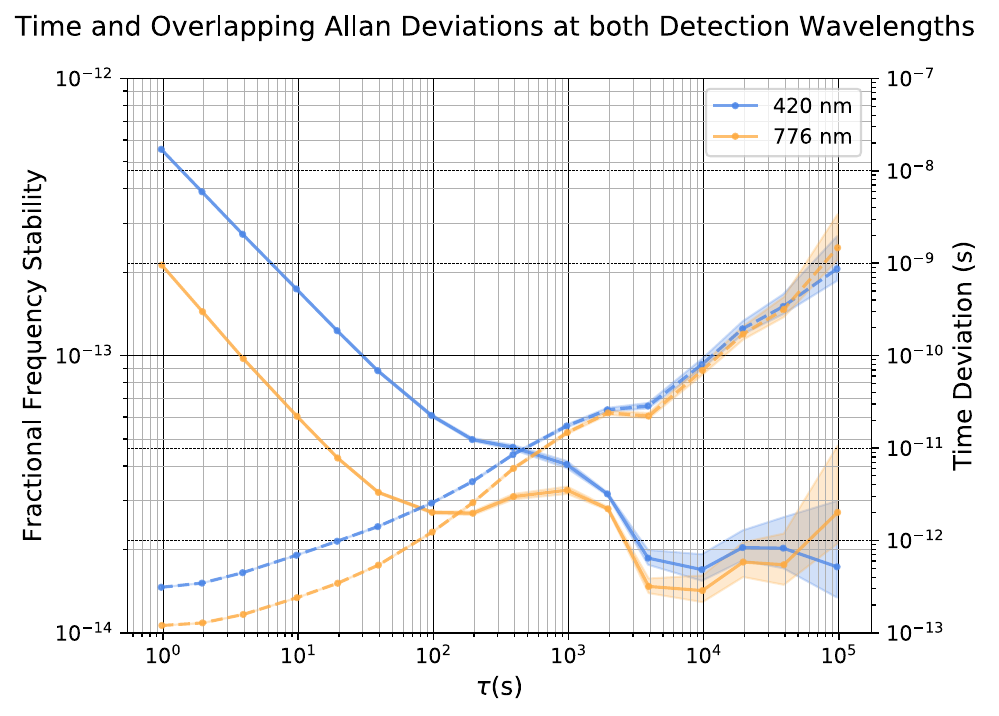}
\caption{\label{BothDevs} Time and Allan deviation of two-photon rubidium clock while detecting 776~nm fluorescence and while detecting 420~nm fluorescence. Frequency instabilities are plotted with solid lines. Integrated timing errors are plotted with dashed lines.}
\end{figure}

The stability of the two-photon clock is measured against an active hydrogen maser, which has a fractional frequency stability of $8\times10^{-14}/\sqrt{\tau(s)}$ . The fractional frequency instabilities of O-RAFS with different detection wavelengths are shown in Fig. \ref{BothDevs}, along with the associated integrated timing errors.

A 2.6$\times$ reduction in short-term instability was observed when detecting at 776~nm compared to detecting at 420~nm ($2.12(1) \times 10^{-13} / \sqrt{\tau(s)}$ and $5.55(1) \times 10^{-13} / \sqrt{\tau(s)}$ at short time-scales, respectively). To check that the clock is not limited by intermodulation noise, the laser power delivered to the cell is temporarily increased such that the short-term fractional frequency instability falls bellow $2 \times 10^{-13} / \sqrt{\tau(s)}$ (not shown). With both detection wavelengths, the clock frequency shows an oscillation at the thousand-second timescale. The source of this oscillation has not been identified for certain, though it is suspected to arise from thermal fluctuations in the lab environment leading to drifts in the control electronics.

\begin{figure}[ht]
\centering
\includegraphics[width=10.3cm]{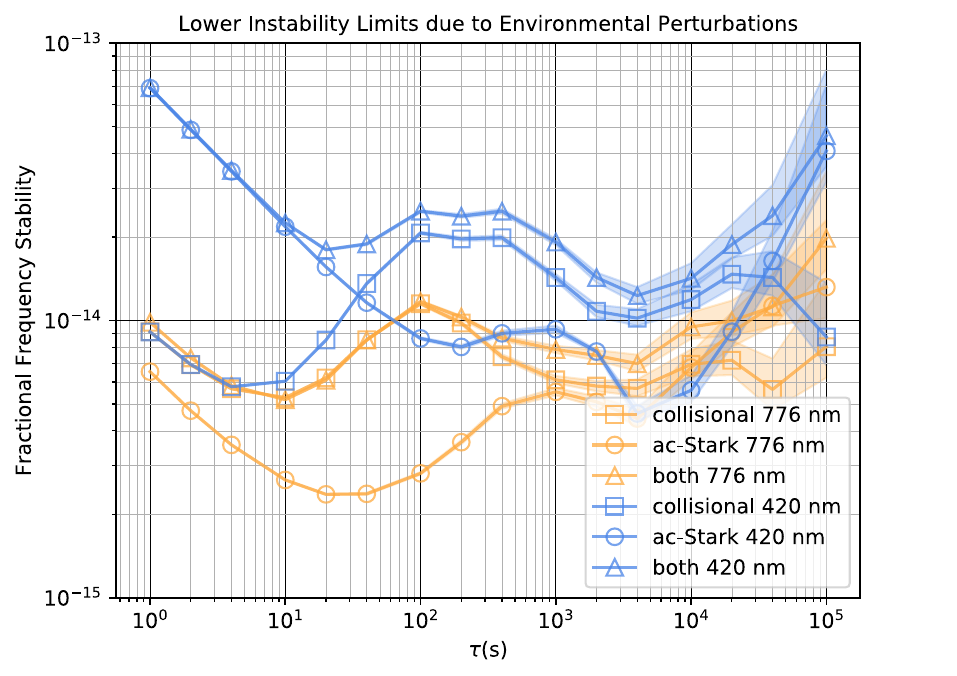}
\caption{\label{EnvironmentalADev} Overlapping Allan deviations of cell temperature and optical power witness data, rescaled to the fractional frequency stability limit they individually impose on the clock based on conversion factors taken from \cite{Martin2018,Lemke2022}. Instabilities from the ac-Stark shift are plotted with circles, instabilities from collisional shifts are plotted with squares, and instabilities from the sum of the shifts are plotted with triangles.
}
\end{figure}

In addition to frequency data, out-of-loop (OOL) sensor data pertaining to environmental parameters that may have affected the clock stability were also recorded, including the vapor cell and baseplate temperature, vapor cell heater and baseplate TEC power, the dc output of the MPPC circuit, and the optical power of the retroreflection coupled back into the optical fiber. Table \ref{EnvTable} summarizes shift rates relating to environmental perturbations and identifies which sensor data were used to estimate the resulting instabilities. Fig. \ref{EnvironmentalADev} displays instabilities from the two largest recorded environmental effects. Other effects were either not recorded during the experiment, such as residual amplitude modulation, or contributed instabilities decades below the relevant noise floor.


\begin{table}[ht]
\begin{center}
\caption{Fractional frequency shift rates due to environmental perturbations and sensors used to estimate contribution to clock instability.}
\label{EnvTable}
\begin{tabular}{ | m{3cm} | m{3cm} | m{6cm} | }
    \hline
    Systematic effect & Shift rate & Sensor data \\
    \hline
    \hline
    Collisional shift & $-1.3\times10^{-30}$~m$^3$ & OOL cell temperature \\
    \hline
    ac-Stark shift & $-8.7 \times 10^{-10}/$~W & Fluorescence and OOL cell temperature\\ 
    \hline
    Zeeman shift & $-4.76 \times 10^{-14}/$~W & Cell heater power\\ 
    \hline
\end{tabular}
\end{center}
\end{table}

\subsection{Collisional Shift}
The collisional frequency shift is taken to be linear with the number density $n$ of the rubidium vapor, which is calculated from the OOL temperature sensor using the model for rubidium vapor pressure $p$ (in Torr) in \cite{SteckRb,Alcock1984}, 

\begin{equation}
    p = 10^{2.881 + 4.312 - 4040/T},
\end{equation}
where $T$ is the temperature of the condensed metal in Kelvin, along with the ideal gas law to obtain

\begin{equation}
\label{nofT}
n = \frac{10^{2.881 + 4.312 - 4040/T}}{kT},
\end{equation}
where k is the Boltzmann constant. The collisional shift at $373$~K is $-7.6 \times 10^{-12}$ \cite{Lemke2022}, which, based on equation \ref{nofT}, corresponds to a collisional shift rate of $-1.3 \times 10^{-30}$~m$^3$.



\subsection{ac-Stark Shift}
The measurement of the optical power of the fiber-coupled retroreflected beam exhibited substantial drift during the experiment, likely due to the sensitivity of free-space to fiber coupling to even small misalignment, and due to drift in the coupling ratio of the fiber splitter used to sample to retroreflected light. This measurement therefore could not be used to estimate the ac-Stark shift. Instead, the output voltage of the MPPC's transimpedance amplifier is used as an OOL sensor to monitor the power of the interrogating beam,  with the caveat that this voltage is also dependent upon the vapor number density. Because this clock probes a weak, nonlinear transition in every velocity class, the dc voltage on the output of the MPPC's transimpedance amplifier is

\begin{equation}
	V \propto n P^2,
\end{equation}
where $n$ is vapor number density and $P$ is the one way optical power (nominally 6~mW). The ac-Stark shift is taken to be linear with optical power and independent of vapor number density. The conversion factor from optical power to the ac-Stark shift is taken from \cite{Martin2018}, however this clock uses a beam waist radius of 0.49~mm instead of the 0.66~mm in \cite{Martin2018}. The proportionality constant between optical power and the ac-Stark shift is rescaled from $-4.8 \times 10^{-10}/$~W to $-8.7 \times 10^{-10}/$~W accordingly.

\subsection{Zeeman Shift}

To estimate the upper boundary on the instability due to the Zeeman shift from the magnetic field of the the resistive heater, the path of the current is modelled as a straight line 6~mm away from the interrogated atoms.

%
Since the magnetic sublevels in this scheme are not resolved, the clock has no first-order Zeeman shift, and the second-order shift must be used instead. Combining the equation for power dissipated across a resistor and the equation for a magnetic field from a line current with the net Zeeman shift rate (quadratic with magnetic field) of $-6.5\times 10^{-3}/$~T$^2$ for the $F=2 \rightarrow F=4$ hyperfine line calculated in \cite{Martin2018}, the fractional frequency shift becomes


\begin{equation}
\frac{\Delta \nu}{\nu} = 6.5\times10^{-3}/\text{T}^2 \times \frac{\mu_0^2}{4 \pi^2 r^2 R} \times P,
\end{equation}
where $\mu_0$ is the permeability of free space, $r = 6$~mm is the distance between the interrogated atoms and the $R = 150$~$\Omega$ resistive heater, and $P$ is the heater power. The shift rate for this clock (linear with heater power) is therefore $-4.76 \times 10^{-14}/$~W, corresponding to a contribution of no more than $2 \times 10^{-16}$ fractional instability at any timescale. However, the heater trace follows a serpentine pattern, generating less magnetic field at the location of the interrogated atoms than a straight wire of the same resistance. Hence, we expect the $2 \times 10^{-16}$ fractional instability to be an overestimate of the actual instability due to the Zeeman shift from the heater field.


\subsection{Other Effects}

Residual amplitude modulation of the interrogating beam was not monitored, and may have contributed to clock instability on long time-scales.

Instabilities due to other effects such as the Zeeman shift from background magnetic fields, the ac-Stark shift from blackbody radiation, and the 2nd-order Doppler shift are taken to be negligible. 
Commercial oscillators of various precision provided






\section{Discussion}

The clock currently shows a 2.6$\times$ reduction in short-term instability in moving from 420~nm detection to 776~nm detection. However, the 776~nm detection method has significant room for improvement, whereas mature, 420~nm-detecting designs such as in \cite{Newman2021,Lemke2022} already collect about as much blue fluorescence as possible. A higher-OD dielectric filter could eliminate the need for the final polarizer, allowing twice as much fluorescence to reach the detector as does now.

We have demonstrated that an MPPC can be used to detect fluorescence in the place of a PMT. This significantly reduces the requirements of the detector's high-voltage power supply. Furthermore, MPPCs have sensitivities ranging much farther into the red and near-infrared than PMTs do, expanding the variety of low-light level processes that can be relied upon in portable quantum sensors. 

Finally, two more points are worth mentioning. First, the 776~nm collection scheme described above allows the fluorescence to expand outward from the optical axis (the axis in line with the interrogating beam) before it reaches a collimating optic of approximately 10~mm diameter clear-aperture, about 2.5 times the diameter of the vapor cell and about 10 times the interrogating beam waist diameter. This expansion is necessary because the fluorescence originates from a non-point source, and so cannot be made uniform in its trajectory without expanding spatially. For some applications this radial expansion of the fluorescence may be permissible, though extremely compact architectures may need to revert to the simpler design of detecting at 420~nm. The advent of a filter with a wider acceptance angle, or a spectrally sharp absorptive filter could alleviate this issue entirely. In an earlier experiment, we aggressively focused the interrogating beam and recollimated it with the collecting lens (effectively placing the cell within a Kepler telescope), thereby exciting atoms only at the focus of the collecting optic, and eliminating the need to let the fluorescence expand. However, the resulting transit-time broadening from thermal atoms rapidly passing through the focus of the beam outweighed the benefits of our increased collection efficiency.

Second, while the atoms emit over nine times as many photons at 776~nm as they do at 420~nm, this optical signal must be converted to an electrical signal, and higher detection efficiencies are currently available at 420~nm than at 776~nm. Because the shot-noise is determined by the number of photons detected, and not just by the number of photons incident on the detector, one must consider the available detector technology when deciding which detection wavelength to use.

\begin{backmatter}
\bmsection{Funding} This work was funded by the Air Force Research Laboratory.

\bmsection{Acknowledgments} We thank Seth E. Erickson for helpful discussion on the optical layout. We thank Nathan D. Lemke, Ben K. Stuhl, and Steve Lipson for careful reading of the manuscript.

\bmsection{Disclaimer}
The views expressed are those of the authors and do not necessarily reflect the official policy or position of the Department of the Air Force, the Department of Defense, or the U.S. government. Contributions to this article by workers at AFRL, an agency of the U.S. government, are not subject to copyright. Any mention of commercial products is for information only, and does not imply endorsement by AFRL.
\\
\textbf{\emph{\small Approved for public release; distribution is unlimited. Public Affairs release approval \#AFRL20236120}}

\bmsection{Disclosures} The authors declare no conflicts of interest.

\smallskip

\bmsection{Data availability} Data underlying the results presented in this paper are not publicly available at this time but may be obtained from the authors upon reasonable request.

\end{backmatter}

\bibliography{sample}

\begin{thebibliography}{10}
\newcommand{\enquote}[1]{``#1''}

\bibitem{Herrmann_2018}
S.~Herrmann, F.~Finke, M.~L\"ulf, \emph{et~al.}, \enquote{Test of the gravitational redshift with galileo satellites in an eccentric orbit,} {\protect\JournalTitle{Phys. Rev. Lett.}} \textbf{121}, 231102 (2018).

\bibitem{Maleki_2005}
L.~Maleki and J.~Prestage, \enquote{Applications of clocks and frequency standards: from the routine to tests of fundamental models,} {\protect\JournalTitle{Metrologia}} \textbf{42}, S145 (2005).

\bibitem{Martin2018}
K.~W. Martin, G.~Phelps, N.~D. Lemke, \emph{et~al.}, \enquote{Compact optical atomic clock based on a two-photon transition in rubidium,} {\protect\JournalTitle{Phys. Rev. Applied}} \textbf{9}, 014019 (2018).

\bibitem{Martin2019}
K.~W. Martin, B.~Stuhl, J.~Eugenio, \emph{et~al.}, \enquote{Frequency shifts due to stark effects on a rubidium two-photon transition,} {\protect\JournalTitle{Phys. Rev. A}} \textbf{100}, 023417 (2019).

\bibitem{Newman2019}
Z.~L. Newman, V.~Maurice, T.~Drake, \emph{et~al.}, \enquote{Architecture for the photonic integration of an optical atomic clock,} {\protect\JournalTitle{Optica}} \textbf{6}, 680--685 (2019).

\bibitem{Newman2021}
Z.~L. Newman, V.~Maurice, C.~Fredrick, \emph{et~al.}, \enquote{High-performance, compact optical standard,} {\protect\JournalTitle{Opt. Lett.}} \textbf{46}, 4702--4705 (2021).

\bibitem{Lemke2022}
N.~D. Lemke, K.~W. Martin, R.~Beard, \emph{et~al.}, \enquote{Measurement of optical rubidium clock frequency spanning 65 days,} {\protect\JournalTitle{Sensors}} \textbf{22} (2022).

\bibitem{Maurice2020}
V.~Maurice, Z.~L. Newman, S.~Dickerson, \emph{et~al.}, \enquote{Miniaturized optical frequency reference for next-generation portable optical clocks,} {\protect\JournalTitle{Opt. Express}} \textbf{28}, 24708--24720 (2020).

\bibitem{Sheng_2008}
D.~Sheng, A.~P\'erez~Galv\'an, and L.~A. Orozco, \enquote{Lifetime measurements of the $5d$ states of rubidium,} {\protect\JournalTitle{Phys. Rev. A}} \textbf{78}, 062506 (2008).

\bibitem{Safronova_2004}
M.~S. Safronova, C.~J. Williams, and C.~W. Clark, \enquote{Relativistic many-body calculations of electric-dipole matrix elements, lifetimes, and polarizabilities in rubidium,} {\protect\JournalTitle{Phys. Rev. A}} \textbf{69}, 022509 (2004).

\bibitem{Safronova_2011}
M.~S. Safronova and U.~I. Safronova, \enquote{Critically evaluated theoretical energies, lifetimes, hyperfine constants, and multipole polarizabilities in $^{87}\mathrm{Rb}$,} {\protect\JournalTitle{Phys. Rev. A}} \textbf{83}, 052508 (2011).

\bibitem{Park_2020}
H.~Park, J.~Tallant, X.~Zhang, \emph{et~al.}, \enquote{171yb+ microwave clock for military and commercial applications,} in \emph{2020 Joint Conference of the IEEE International Frequency Control Symposium and International Symposium on Applications of Ferroelectrics (IFCS-ISAF),}  (2020), pp. 1--4.

\bibitem{Xin_2022}
N.~C. Xin, H.~R. Qin, S.~N. Miao, \emph{et~al.}, \enquote{Laser-cooled 171yb$+$ microwave frequency standard with a short-term frequency instability of 8.5 {\texttimes} 10{\textminus}13/\&\#x221a;\&\#x3c4;,} {\protect\JournalTitle{Opt. Express}} \textbf{30}, 14574--14585 (2022).

\bibitem{Schuldt_2021}
T.~Schuldt, M.~Gohlke, M.~Oswald, \emph{et~al.}, \enquote{Optical clock technologies for global navigation satellite systems,} {\protect\JournalTitle{GPS Solutions}} \textbf{25} (2021).

\bibitem{Doeringshoff_2019}
K.~D\"oringshoff, F.~B. Gutsch, V.~Schkolnik, \emph{et~al.}, \enquote{Iodine frequency reference on a sounding rocket,} {\protect\JournalTitle{Phys. Rev. Applied}} \textbf{11}, 054068 (2019).

\bibitem{Roslund2023}
J.~D. Roslund, A.~Cingöz, W.~D. Lunden, \emph{et~al.}, \enquote{Optical clocks at sea,}  (2023). ArXiv:2308.12457 [physics.atom-ph].

\bibitem{IFCSPoster}
R.~Beard, K.~W. Martin, J.~D. Elgin, \emph{et~al.}, \enquote{Poster: Detection of 776 nm fluorescence in two-photon rubidium clock with multi-pixel photon counter,} in \emph{International Frequency Control Symposium,}  (Toyama, Japan, 2023).

\bibitem{Hassasin2023}
K.~Hassanin, P.~Federsel, F.~Karlewski, and C.~Zimmermann, \enquote{$5s\text{\ensuremath{-}}5d$ two-photon transition in rubidium vapor at high densities,} {\protect\JournalTitle{Phys. Rev. A}} \textbf{107}, 043104 (2023).

\bibitem{UDportal}
P.~Barakhshan, A.~Marrs, A.~Bhosale, \emph{et~al.}, Portal for High-Precision Atomic Data and Computation (version 2.0). University of Delaware, Newark, DE, USA. URL: https://www.udel.edu/atom/[February 2022].

\bibitem{Setzer2021}
W.~J. Setzer, M.~Ivory, O.~Slobodyan, \emph{et~al.}, \enquote{Fluorescence detection of a trapped ion with a monolithically integrated single-photon-counting avalanche diode,} {\protect\JournalTitle{Applied Physics Letters}} \textbf{119}, 154002 (2021).

\bibitem{SteckRb}
D.~A. Steck, \enquote{"rubidium 87 d line data," available online at http://steck.us/alkalidata (revision 2.2.2, 9 july 2021),} .

\bibitem{Alcock1984}
M.~K.~H. C.~B.~Alcock, V. P.~Itkin, \enquote{Vapor pressure equations for the metallic elements: 298-2500 k,} {\protect\JournalTitle{Canadian Metallurgical Quarterly}} \textbf{309} (1984).

\end{thebibliography}






\end{document}